\documentclass[twocolumn,showpacs,preprintnumbers,amsmath,amssymb]{revtex4}

\usepackage{graphicx}

\newcommand{\beq}{\begin{equation}}
\newcommand{\eeq}{\end{equation}}

\newcommand{\beqa}{\begin{eqnarray}}
\newcommand{\eeqa}{\end{eqnarray}}

\font\blackboard=msbm10 at 10pt
\font\blackboards=msbm7
\font\blackboardss=msbm5
\newfam\black
\textfont\black=\blackboard
\scriptfont\black=\blackboards
\scriptscriptfont\black=\blackboardss
\def\mathbb#1{{\fam\black\relax#1}}

\newcommand{\BZ}{{\mathbb Z}}

\newcommand{\CL}{{\cal L}}

\renewcommand{\bar}{\overline}

\def\half{{\textstyle{1\over 2}}}
\def\pa{\partial}

\newcommand{\Tr}{{\rm Tr}}

\begin{document}

\preprint{hep-ph/0306223}
\title{Implications of Local Chiral Symmetry Breaking}
\author{HoSeong La}
\affiliation{
Department of Physics and Astronomy,\\
University of Mississippi, University, MS 38677, USA\\
{\tt hsla@phy.olemiss.edu}}
\date{June 23, 2003}

\begin{abstract}
The spontaneous symmetry breaking of a local chiral symmetry to its diagonal
vector symmetry naturally realizes a complete geometrical structure more
general than that of Yang-Mills (YM) theory, rather similar to that of
gravity. A good example is the Quantum Chromodynamics (QCD) with respect to
the Chiral Color model. Also, a new anomaly-free particle content for
a Chiral Color model is introduced: the Chiral Color can be realized
without introducing whole new generations of quarks and leptons, but by
simply enlarging each generation with new exotic fermions.
\end{abstract}


\maketitle

\setcounter{footnote}{0}
\setcounter{page}{1}
\setcounter{section}{0}
\setcounter{subsection}{0}
\setcounter{subsubsection}{0}


\section{Introduction}

One of the most important challenges in the high energy physics today is to
find out what lies beyond the standard model. Since the standard model is
based on gauge theory, one of the simplest attempts is to consider
extending the gauge symmetry in four dimensions.
Models in this line of extension includes,
most prominently, the grand unification models\cite{GG,PS}
and, more modestly, the left-right symmetric model\cite{PS,MP} and
(extended) technicolor models\cite{tc}, etc.

The Chiral Color\cite{fg} is another model with an extended gauge symmetry
in this spirit. It was proposed as a generalization of QCD such that
QCD, which is vector-like, may be a result of spontaneous symmetry breaking
of the Chiral Color.
The gauge group is two copies of SU(3) with a global $\BZ_2$ symmetry,
i.e. the left-right symmetry, to start with (other models having two
copies of one or more components of the standard model gauge group can
be found in \cite{topcolor,KL,AS,BKP}):
\beq
G_{\rm CC}\equiv {\rm SU}(3)_L\times {\rm SU}(3)_R.
\eeq
Chiral Symmetry in QCD is often considered as a global flavor symmetry, whose
symmetry breaking can explain the origin of (light) quark masses. Note that,
on the contrary, the symmetry $G_{\rm CC}$ we consider in this letter is
a local chiral symmetry.

Generically, the gauge fields of ${\rm SU}(N)_1\times {\rm SU}(N)_2$
can be mixed such that
\beq
\label{e1}
\left( B_\mu \atop u_\mu\right)
\equiv
\begin{pmatrix}
\cos\theta & \sin\theta \\
-\sin\theta & \cos\theta
\end{pmatrix}
\left( A_\mu^{(1)} \atop A_\mu^{(2)} \right),
\eeq
where the coupling constants are related as
\beq
\label{e2}
{g}\equiv g_1\cos\theta =g_2\sin\theta
\eeq
and $g$ is the coupling constant of the unbroken SU$(N)$.
In our notation we will use $B_\mu$ to denote the gauge fields of the
unbroken vector-like SU$(N)$.
There are two different types of embeddings:
First, $g_1=g_2=\sqrt{2}g$, which corresponds to the Chiral Color
with the left-right symmetry.
Second, without imposing  the left-right symmetry, let one of the coupling
constants, say, $g_1\to \infty$ and the other to be $g=g_2$,
without loss of generality. This corresponds to the Topcolor
(when $N=3$ and both are vector-like)\cite{topcolor,tc}.

In the Chiral Color case,  $G_{\rm CC}$ must be broken spontaneously
to ${\rm SU}(3)_{\rm C}$
\footnote{Such a Chiral Color structure first showed up in one of
Pati-Salam models\cite{PS}. },
leaving eight massive axial vector bosons called axigluons.
The search so far has ruled out axigluons lighter than
1 TeV\cite{axiexp}. Nevertheless, it is still a very interesting idea
since there is no {\it a priori} reason axigluons must be light.

In this letter we provide other reasons why we should further pursue
the Chiral Color. These are two-folds:
First, the mystery of the YM geometry can be clarified and it brings up 
a new geometrical structure for a gauge theory similar to that of gravity.
Second, the Chiral Color can be realized by
enlarging the content of each generation without introducing new fermion
generations beyond that of the standard model. This is done by minimally
introducing new exotic fermions in each generation.
The latter indicates this new Chiral Color model is not a technicolor
model in which technifermions are introduced as separate generations, and
the former may provide a new clue toward nonperturbative aspects
of QCD as well as a new way of relating the gauge theory and gravity.

\section{Geometry of Local Chiral Symmetry}

Under $G_{\rm CC}$, the left-right symmetry implies that the roles of
two sets of gauge fields are not particularly different. However,
upon symmetry breaking $G_{\rm CC}$ to SU(3)$_{\rm C}$,
the difference between $B_\mu$ and $u_\mu$ clearly
emerges. $B_\mu$ still transforms as a gauge field under SU(3)$_{\rm C}$, but
$u_\mu$ now transforms covariantly:
\beqa
B_\mu &\to& U^{-1}(B_\mu +\pa_\mu)U, \cr
u_\mu &\to& U^{-1} u_\mu U.
\eeqa
Under the left-right symmetry, $B_\mu$ is even and $u_\mu$ is odd.
In the sense that $u_\mu$ transforms covariantly, the axigluon $u_\mu$ takes
the role of ``vector matter'' under SU(3)$_{\rm C}$. One may be tempted to
introduce this type of vector matter by hand to QCD.
But, the massless $u_\mu$ case is just a field redefinition of unbroken
$G_{\rm CC}$. Since the mass term of $u_\mu$ is  SU(3)$_{\rm C}$ invariant,
one may add the mass term by hand to distinguish it. 
Unfortunately, this massive case turns out to be non-renormalizable.
This confirms that, as far as vector fields are concerned,
the gauge invariance alone is not good enough to ensure
the renormalizability. As is well known, the mass of a gauge boson must
come from a Higgs mechanism to be renormalizable.
In our case, this argument is extended further that the mass of a covariant
vector field cannot be introduced by hand to be renormalizable.

The gauge invariant coupling of $u_\mu$ to $B_\mu$ can be elegantly expressed
by defining\cite{monoloop}
\beq
\label{etorsion}
H_{\mu\nu}^a \equiv \pa_\mu u_\nu^a -\pa_\nu u_\mu^a
+f^{abc}\left(B_\mu^b u_\nu^c-B_\nu^b u_\mu^c\right).
\eeq
Note that $H_{\mu\nu}$ transforms covariantly under SU(3)$_{\rm C}$. 
Then the gauge field part of the lagrangian can be rewriten as follows:
\beqa
\label{e10}
\CL_{\rm CC}&=& -\half\Tr \left(F_{\mu\nu}^{(1)}\right)^2
-\half\Tr \left(F_{\mu\nu}^{(2)}\right)^2\cr
&=& -\half \Tr \left( F_{\mu\nu}+[u_\mu, u_\nu]\right)^2
-\half \Tr \left(H_{\mu\nu}\right)^2.
\eeqa
Due to rather complicated couplings between axigluons and gauge fields,
the first term is expressed in a somewhat unfamiliar way.
Nevertheless, the Bogomol'nyi-Prasad-Sommerfeld (BPS) limit still contains
that of SU(3)$_{\rm C}$ case. In fact, the instanton solution of
SU(3)$_{\rm C}$ case corresponds to the common instanton solution
of SU(3)$_L$ and SU(3)$_R$ when the sizes and instanton numbers of these
instantons are identical and $u_\mu^a=0$,
in accordance with the left-right symmetry of the Chiral Color.
Upon Higgs Mechanism, the mass term of $u_\mu$ is to be added.

As is well known, the usual geometry of YM theory only involves a (spin)
connection, $B_\mu^a$, and a curvature, $F_{\mu\nu}^a$. Here we have
$u_\mu^a$ in addition. Looking at the structure of Eq.(\ref{etorsion}),
we can identify $H_{\mu\nu}^a$ as a torsion and $u_\mu^a$ as vielbeins.

Although it has been clearly emphasized by Yang\cite{WY,Yang:1979tx} that
the geometry of YM theory is based on a principal (fiber) bundle, in which
the dynamical variable is a gauge field and it behaves likes a connection 
on the group manifold. On the contrary, the geometry of General Relativity (GR)
is based on a tangent bundle, in which the metric is a fundamental dynamical
variable. Nevertheless, one cannot help but wondering why
the analog of a metric or a vielbein is missing in the geometry of YM theory.
In this letter, we have presented the answer to resolve
this mystery: these extra data are present if we consider a vector gauge
theory inherited from a local chiral symmetry and its breaking.

The appearance of the vielbeins in the geometry of YM theory was first alluded
in \cite{monoloop} to construct the virtual monopole geometry associated
with BPS monopole solution in the SU(2) YM theory. The structure is a
generalization of covariant variables introduced by Johnson and
Haagensen\cite{HJ,HJL}, in which a constraint equation related the SU(2) gauge
fields to these covariant variables. In \cite{HJ,HJL}, however,
these covariant variables are not field variables that show up in the
Lagrangian. In the Chiral Color model these gauge covariant fields are indeed
present in the Lagrangian and take the role of the vielbeins with respect
to the unbroken local gauge symmetry, as presented here.
In the limit $H_{\mu\nu}^a = 0$, which corresponds to the torsion-free
condition in this context, the gauge field $B_{\mu}^a$ can be expressed
in terms of axial gauge field $u_\mu^a$.
However, this is not necessarily a desired
structure in this model. Perhaps, there might be interesting phenomena
involving nonvanishing torsion in the line of the idea presented in
\cite{monoloop}, in which three different phases of QCD (Coulomb, confinement
and, suspectedly,  oblique confinement) are associated
with three different asymptotic behaviors of the torsion.
This resolves the difficulty
of relaxing the torsion-free condition in \cite{HJ, HJL} since in the
Chiral Color the torsion is unambiguously present
because the axial gauge field $u_\mu$ is an independent dynamical variable.

\section{Chiral Color with Enlarged Generations}

This raises a question if Nature actually likes the proposed structure.
If there is no realistic model what so ever, it could indicate that 
Nature prefers the geometric data of YM theory being fundamentally 
different from that of gravity. 
So It all depends on the existence of a realistic Chiral Color model.
The answer to this quest is affirmative and the model we present here is
without unnecessary fermion generations beyond the standard model
yet containing interesting particle content.

Under the extended gauge symmetry
\beq
G=
{\rm SU}(3)_L \times {\rm SU}(3)_R
\times {\rm SU}(2)_L \times {\rm U}(1)_Y,
\eeq
the quarks and leptons in each generation transform as
\beqa
&(3,1,2, 1/6)_L,\  (1,3,1,2/3)_R,\ (1,3,1,-1/3)_R, \cr &\cr
&(1,1,2,-1/2)_L,\ (1,1,1,-1)_R.
\eeqa
Since the charge assignment is the same as that of the standard model, the
${\rm SU}(2)_L\times {\rm U}(1)_Y$ anomalies cancel. In the Chiral Color
case there is an additional anomaly due to the chiral nature of SU(3)'s.
There are various ways of canceling the SU(3) anomaly by introducing
additional fermions, as  presented in \cite{fg}. But all of them employs
additional generations, which inevitably introduce new lepton flavors.
New lepton flavors are unwelcome unless we can make sure the accompanying
neutrinos are sterile.

In this letter, we will cancel the SU(3) anomaly without introducing
a new generation hence avoiding unwanted lepton flavors, but by enlarging
the generation itself. Since there is no
{\it a priori} reason that at higher energy scales the only matters available
should be still quarks and leptons, one can argue that there could be
additional siblings to quarks and leptons in each generation,
massive and exotic enough to be unobservable at the standard model energy
scale. The choice we make to cancel the SU(3) anomaly due to quarks is
\beq
(\bar{3}, 1, 2, 0)_L,\ (1,\bar{3},1,1/2)_R,\ (1,\bar{3}, 1, -1/2)_R.
\eeq
The hypercharges are assigned as shown such that there are no new
${\rm SU}(2)_L\times {\rm U}(1)_Y$ anomalies generated by these extra
fermions. As a result, these extra fermions carry half-integer electric
charges. All anomalies are canceled in each generation separately as in the
standard model.

These new exotic fermions are very much like (anti-) quarks except carrying
different electric charges. Due to the electric charge conservation they
do not directly couple to quarks and leptons. 
Most likely, these exotic fermions are characteristically more massive than 
quarks. They go through electroweak processes.
Neutral currents can produce exotic fermions in pairs. These signals look
very much alike quark pair production. The distinction will be made 
by measuring the electric charge. Also high energy quarks and leptons 
can produce these exotic fermions via a W boson process. A good place to look 
for such a signal is the atmospheric interaction with ultra high energy 
cosmic neutrinos.

The presence of hybrid bound states between these exotic fermions and quarks 
is highly undesirable, although not necessarily ruled out experimentally.
Note that the choice of assigning these exotic fermions as $\bar{3}$
forbids at least scalar hybrid bound states because the color singlet
combination between a quark and an exotic fermion cannot be a Lorentz scalar.
Other hybrid states may exist and carry fractional charges in multiples of
${1\over 6}e$. Among exotic fermions,
the color singlet fermion-antifermion, or three-fermion bound states
can be formed and the three-fermion states carry half-odd-integer
electric charges. The observation of such a fractionally charged states
will tell us if Nature likes such a new structure in addition to the 
observation of axigluons.

Unless there
is other hidden structure in the more extended context, e.g. the existence
of a grand unified model incorporating these exotic fermions, the first
generation of them must be stable. Certainly, at this moment we do not have
any evidence for their existence in the macroscopic world in which only
the first generation of quarks and leptons are clearly present\cite{DCB}.
However, the search for a fractional electric charge particle has been still
actively going on by a team led by Perl\cite{perl}. 
One of the search is looking at bulk matter from outside the solar system 
because such a particle might have been produced in the early universe. 
Due to lack of direct coupling to quarks, it is quite possible these 
exotic fermions might have been decoupled very early on at the electroweak 
scale.

This also rehashes the elusive question of the electric charge quantization.
All observable particles in Nature so far carry integer multiplication of the
elementary charge and one justification for this electric charge
quantization is the existence of a Dirac magnetic monopole.
However, we have not found any evidence of the existence of a magnetic
monopole, hence the electric charge quantization is still an open question.
Discovery of a fractionally charged particle may
change our current preference on this issue: perhaps electric charge is not
quantized by integer multiplication, but by one-sixth-integer multiplication.
As a matter of fact, even the existence of a magnetic monopole does not
necessarily dictate the integer multiplication of electric charge
quantization. For example, if a nonabelian magnetic monopole is produced 
as a $\BZ_2$ symmetric vortex,
which is the case of ${\rm SU}(2)\rightarrow {\rm U}(1)$ breaking by an adjoint
higgs, the electric charge is quantized by half-integer
multiplication\cite{thooftmono}. Also, in the string-inspired models, $\BZ_N$
discrete symmetry which commonly appears in the orbifold compactification
leads to $1/N$ fractional charges\cite{Ant}. Perhaps, there may be a hidden
stucture with $\BZ_6$ discrete symmetry at the Chiral Color scale, which 
can explain even why quarks carry fractional charges.

In the original Chiral
Color model, the Chiral Color symmetry breaking scale was assumed to be
the same as the electroweak symmetry breaking scale. This can certainly be
relaxed and we can push the Chiral Color symmetry breaking scale higher than
the electroweak symmetry breaking scale.
So the symmetry breaking of $G_{\rm CC}$ to SU(3)$_{\rm C}$ can be achieved,
in the simplest case, by a scalar multiplet transforming as $(3,\bar{3},1,1)$.
Although the Chiral Color scale can be, in principle, anywhere, there is
one desirable scale we can speculate.
Suppose there were a relationship between scales
such that the square of the electroweak scale is directly related to the
multiplication of the Chiral Color scale and the QCD scale. This will put
the Chiral Color scale at about $10\sim 100$ TeV.
This will be the most interesting Chiral Color scale because it is less
{\it ad hoc} than having three totally unrelated scales.
Furthermore, this Chiral Color scale is certainly reachable in the near
future at the LHC (Large Hadron Collider), hence we may begin to see the
Chiral Color signals.

Note that unlike in the technicolor models, these new fermions do not
exclusively interact with axial vector combination of $G_{\rm CC}$. This also
distinguishes our model from technicolor models.

\section{Remarks}

Finally, a few remarks on further works to investigate the implications of
the structure presented in this letter are in order.

This new geometry of YM theory can be a link between QCD and QCD string,
perhaps a detail study may shed new light on nonperturbative aspects of QCD.
The geometrical structure presented in this letter also opens up a new way
of thinking about the gravity as a gauge theory since it has all the
necessary geometrical data. The spacetime metric can
be identified as an induced metric
\beq
g_{\mu\nu}=\xi_{ab}u_\mu^a u_\nu^b,
\eeq
where $\xi_{ab}$ is the Cartan metric of the gauge group.
This in turn can be solved for spacetime vielbeins as
\beq
g_{\mu\nu}=\eta_{AB}e_\mu^A e_\nu^B
\eeq
where $A,B$ are indices for the local Lorentz symmetry and
$\eta_{AB}$ is the Minkowski metric.
It will be a good mission to look for a gauge group for which gravity can
be realized this way, and the search is in progress.

Our proposal of extending the standard model using the Chiral Color opens
up many interesting possibilities we can further investigate. Not only looking
for axigluons, but we can also search for the possibility of fractionally
charged three-fermion states to see if Nature really respects the
charge quantization rule beyond the standard model. There are also varieties of
signals involving the new exotic fermions. It should certainly provide
new opportunities for future accelerators and particle astrophysics.

There are other new aspects we can study in this context.
Quarks now interact not only with gluons but with axigluons.
Being massive, the axigluons will not change the long distance behavior of
quarks. However, the short distance behavior of quarks will be modified.
The short distance interquark potential will not be just coulombic, but
may have a Yukawa type correction.

The mass matrix of the new exotic fermions should have an analogous structure
as the Cabibbo-Kobayashi-Maskawa matrix of quarks,
hence possibly induces a CP-violation in the exotic fermion processes.
This new CP-violation can contribute to the matter-antimatter asymmetry,
helping the current discrepancy between the observed
baryon asymmetry and theoretical estimations. We can speculate that some
of the matter identified as baryons could be matter formed by these new
exotic fermions since we never considered this possibility so far.

We believe this letter presents a good motivation to look into the role
of local chiral symmetry and its breaking seriously, 
not just global chiral symmetry we have studied so far. 
The Chiral Color model has such fascinating
structures, Nature may provide an evidence of its relevance in the near future.
Further progress will be presented elsewhere.

\acknowledgements{The author thanks S. Coleman and S. Glashow
for conversations in the early stage of this work.}

\end{document}